\documentclass[aps,prx,twocolumn]{revtex4-1} 
\usepackage{graphicx}
\usepackage{amsmath,amsfonts,amssymb}
\usepackage{bm}
\usepackage{color}

\usepackage{xcolor}

\usepackage{ulem}

\usepackage{soul}

\begin{document}

\title{Inertial self-propelled particles}

\author{Lorenzo Caprini}
\affiliation{Scuola di Scienze e Tecnologie, Universit\`a di Camerino,  
Via Madonna delle Carceri, I-62032, Camerino, Italy}
  

\author{Umberto Marini Bettolo Marconi}
\affiliation{Scuola di Scienze e Tecnologie, Universit\`a di Camerino,  
Via Madonna delle Carceri, I-62032, Camerino, Italy}

\begin{abstract}
We study a self-propelled particle moving in a solvent with the active Ornstein Uhlenbeck dynamics in the underdamped regime to evaluate the influence of the inertia.
We focus on the properties of potential-free and harmonically confined underdamped active particles, studying how the single-particle trajectories modify for different values of the drag coefficient. In both cases, we solve the dynamics in terms of correlation matrices and steady-state probability distribution functions revealing the explicit correlations between velocity and active force.
We also evaluate the influence of the inertia on the time-dependent properties of the system, discussing the mean square displacement and the time-correlations of particle positions and velocities.
Particular attention is devoted to the study of the Virial active pressure unveiling the role of the inertia on this observable. 
\end{abstract}

\maketitle

\section{Introduction}

Self-propelled systems, that include bacteria, algae, cell monolayers, artificial microswimmers, and micro-organisms, perform their active motion in a solvent where they experience large frictional forces, in general, much stronger than inertial forces~\cite{bechinger2016active, marchetti2013hydrodynamics}.
It is then safe, when the particle sizes are on the scale of nanometers or microns, to neglect the effects of their acceleration.
These systems are usually modeled by means of overdamped Brownian dynamics with standard forces, induced by pairwise interactions and/or external potentials, but are called ``active'' because of the presence of a directed force, that changes stochastically with a typical persistence time~\cite{gompper20202020, ma2020dynamic}.

While most studies focus on the overdamped limit of Brownian dynamics, the influence of the particle inertia on the properties of active matter has been the subject of some recent studies~\cite{lowen2020inertial}.
Several types of robots behaving as self-propelled particles have been fabricated at macroscopic length scales.
Examples are vibro-robots self-propelling through tilted elastic legs~\cite{scholz2018inertial, deblais2018boundaries}, Hexbug crawlers~\cite{dauchot2019dynamics}, camphor surfers, which glide at the air-fluid interface~\cite{bourgoin2020kolmogorovian, leoni2020surfing}, or rotating robots~\cite{scholz2018rotating}.
Moreover, new microfluidic devices forming a new class of self-propelling particles have been realized employing semiconductor diodes whose energy is provided by a global external a.c. electric field.
In general, millimeter-sized active particles moving in low viscosity media are deeply affected by inertial forces
since the role of the fluidic drag is reduced.

From a theoretical point of view, we may expect that, in the absence of external forces, inertia induces a longer time delay for particles to change their speed, increases the persistence of trajectories and increases the particle mobility.
Moreover, evaluating the interplay between inertia and self-propulsion may have consequences on the particle trajectory that we are going to reveal both in confinement-free and harmonically trapped particles.
In addition, since harmonically confined passive particles display oscillating time correlation functions in the underdamped regime, we are going to show how this scenario is modified by the presence of the self-propulsion showing also a comparison with the well-known results in the overdamped active regime.
Finally, the second-order equation of motion for underdamped Langevin dynamics is needed to describe more complex mechanisms such as the joint effect of self-propulsion and magnetic fields~\cite{vuijk2020lorentz}.

The understanding of these aspects requires a description taking into account the
acceleration of the particles, in contrast with the one commonly employed to describe self-propelled systems.
As recent studies have shown, inertia affects many properties of active particles, such as their pressure~\cite{joyeux2016pressure, takatori2017inertial, fily2017mechanical, sandoval2020pressure, gutierrez2020inertial}, transport properties~\cite{ai2017transport, zhu2018transport}, the stochastic energetics~\cite{shankar2018hidden} and, even, anomalous responses to boundary driving~\cite{wagner2019response}.
Besides, inertial forces play an important role also at the collective level: i) affecting the clustering typical of active matter and, in particular, suppressing the phase-coexistence~\cite{suma2014motility, manacorda2017lattice, petrelli2018active, mandal2019motility, dai2020phase} and changing several features of the transition~\cite{su2020inertial} ii) modifying some properties of dense phases of active matter~\cite{arold2020active}, such as the active temperature in the homogeneous~\cite{de2020phase} and inhomogeneous phases~\cite{petrelli2020effective}.

 
The article is structured as follows: in Sec.~\ref{Sec:Model}, we introduce the model used to describe a self-propelled particle in the underdamped regime unconfined or confined through an external potential. 
In Sec.~\ref{Sec:free}, the potential-free particle is investigated while, in Sec.~\ref{Sec:Harmonic}, we evaluate the case of harmonic confining potential.
In both cases, we discuss single-particle trajectories, steady-state properties, such as correlation matrix and steady-state probability distributions, and time-dependent properties, i.e. mean square displacement and temporal position and/or velocity correlations. Sec.~\ref{Sec:Virial} discuss the virial pressure while some conclusive remarks and discussions are reported in the final section.

\section{Model}
\label{Sec:Model}

We study the dynamics of a self-propelled particle evolving with the underdamped version of the Active Ornstein-Uhlenbeck particle (AOUP) model, to understand the role of the inertia on the particle dynamics.
The overdamped AOUP is a well-known active matter model~\cite{maggi2017memory, berthier2017active, caprini2018activeescape, wittmann2018effective, dabelow2019irreversibility, berthier2019glassy, woillez2020nonlocal, martin2020statistical} that captures many characteristic properties of self-propelled particles, including the accumulation near rigid boundaries~\cite{caprini2019activechiral, maggi2015multidimensional, wittmann2016active} or obstacles and non-equilibrium phase coexistence~\cite{fodor2016far, maggi2020universality}.
This model has been successfully compared to another popular model in the active matter community, the Active Brownian particle model (ABP), through studies that shed light on their connections~\cite{caprini2019comparative, das2018confined}.
Despite the numerous applications of the ABP model, we remark the theoretical advantages in the use of the AOUP stemming from its simplicity. Starting from the AOUP, several predictions or approximations for the probability distribution~\cite{fodor2016far, wittmann2017effective, bonilla2019active}, pressures~\cite{caprini2018active, wittmann2019pressure} and surface currents near boundaries~\cite{caprini2018active, caprini2019activechiral} have been derived.
More recently, AOUP has been employed to derive the analytical expression for the spatial correlation of the velocity spontaneously appearing in active hexatic and solid phases~\cite{caprini2020hidden, caprini2020time}.

The dynamics of the underdamped AOUP model for a particle of mass $m$ is governed by the following stochastic equations:
\begin{subequations}
\label{eq:motion}
\begin{align}
\dot{\mathbf{x}}&=\mathbf{v} \,,\\
\dot{\mathbf{v}}& = - \gamma \mathbf{v} +\frac{ \mathbf{F}}{m} + \frac{\mathbf{f}_a}{m} + \sqrt{2 \gamma \frac{T}{m}} \boldsymbol{\eta} \,,
\end{align}
\end{subequations}
being $\mathbf{x}$ and $\mathbf{v}$ the particle position and velocity, respectively.
$\gamma$ and $T$ are the drag and temperature of the solvent, that satisfy the Einstein relation, namely $\gamma D_t= T/m$, with the thermal diffusion coefficient, $D_t$.
The term $\boldsymbol{\eta}$ is a white noise vector with zero average and unit variance which accounts for the collisions between the self-propelled particle and the particles of the solvent, such that $\langle \eta_i(t) \eta_j(t')\rangle=\delta(t-t')$.
As for equilibrium colloids, the solvent exerts a Stokes force proportional to $\mathbf{v}$.
In the case of many active colloids and bacteria~\cite{bechinger2016active}, the thermal diffusivity is often negligible with respect to the effective diffusivity associated with the active force,  but some recent experiments have shown that in some situations it must be accounted for~\cite{dauchot2019dynamics}.
The force contribution is accounted through the term $\mathbf{F}$ which represents the force due to an external potential determined by $\mathbf{F} = - \nabla U$.
The term $\mathbf{f}_i^a$ is introduced to model the so-called self-propulsion, i.e. a complex force with chemical or mechanical origins whose details are usually dismissed at this level of description.
This force pushes the system far from equilibrium, storing energy from the environment and converting it into directed  motion, and usually guarantees a persistent motion in one direction at least for short times (i.e. smaller than the persistence time).
According to the AOUP model, $\mathbf{f}_a$ evolves through an Ornstein-Uhlenbeck process:
\begin{equation}\label{eq:motion2}
\tau\dot{\mathbf{f}}_a = -\mathbf{ f}_a + f_0\sqrt{2 \tau} \boldsymbol{\xi} \,,
\end{equation}
where $\boldsymbol{\xi}$ is a white noise vector with zero average and unit variance. 
The parameter $\tau$ is the persistence time of the active force, determining the typical rate change of $\mathbf{f}_a$, while $f_0$ is the average value taken by the modulus of the active force, fixing the swim velocity induced by the self-propulsion: 
$$
v_0=\frac{f_0}{m\gamma} \,.
$$
The larger is the friction (or mass) the larger is the amplitude of the self-propulsion needed to induce the same $v_0$.
Roughly for a time $\tau$, the self-propulsion does not change direction and provides a constant force contribution, i.e. a swim velocity persisting in a random direction for a time $\tau$.

We recall that for generic potentials even in the overdamped case the form of the probability distribution and of the relevant correlation functions are not known~\cite{martin2020statistical}.
The only exceptions are the potential-free and the harmonically confined AOUP cases, where the linearity of the dynamics allows us to find the explicit solution in terms of correlations and steady-state probability distribution functions~\cite{szamel2014self, caprini2018linear}.

\section{Free Particles}
\label{Sec:free}

\begin{figure}[t]
\centering
\includegraphics[width=0.99\columnwidth,keepaspectratio]
{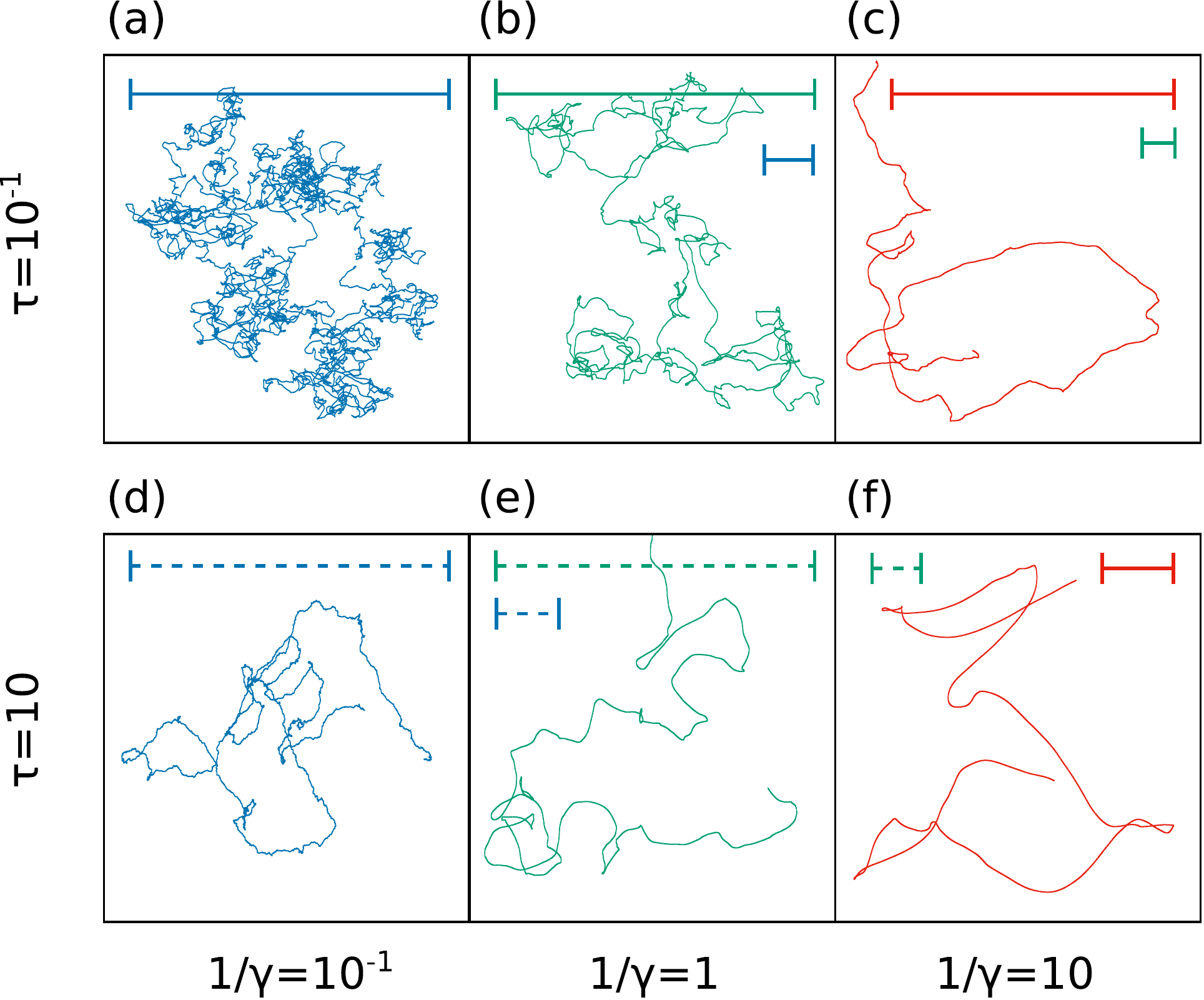}
\caption{\label{fig:trajfree}. 
Single-particle trajectories in the plane $xy$ evolving up to the same final time $\mathcal{T}=10^2$.
Panels (a), (b) and (c) (upper panels) are obtained with $\tau=10^{-1}$ while panels (d), (e) and (f) (bottom panels) with $\tau=10$.
Instead the couples [(a),(c)], [(b), (d)] and [(c), (f)] have been realized with $1/\gamma=10^{-1}, 1, 10$, respectively.
Trajectories are colored according to the value of $1/\gamma$.
The horizontal straight lines drawn in some panels represent the average lengths traveled within the time $\mathcal{T}$ by the configurations of the other panels and follow the same notation of colors. Solid and dashed eye-guides correspond to configurations realized with $\tau=10^{-1}, 10$, respectively. 
The trajectories are obtained with $T=10^{-3}$ and $f_0=1$.
}
\end{figure}

In the absence of external potentials, the dynamics \eqref{eq:motion} reduces to:
\begin{subequations}
\label{eq:motion_free}
\begin{align}
\dot{\mathbf{x}}&=\mathbf{v} \,,\\
\label{eq:v_free}
\dot{\mathbf{v}}& = - \gamma \mathbf{v}  + \frac{\mathbf{f}_a}{m} + \sqrt{2 \gamma \frac{T}{m}} \boldsymbol{\eta} \,,\\
\label{eq:fa_free}
\tau\dot{\mathbf{f}}_a &= -\mathbf{ f}_a + f_0\sqrt{2 \tau} \boldsymbol{\xi} \,,
\end{align}
\end{subequations}
Being confinement-free, the particle position experiences a diffusive motion for long times.
In this simple case, the dynamics is governed by two typical times, the inertial time, $1/\gamma$, and the persistence time, $\tau$, and there are no other mechanisms that could affect the dynamics.

Fig.~\ref{fig:trajfree} shows several two-dimensional single-particle trajectories realized with different settings of the parameters, varying $\gamma$ and $\tau$, keeping fixed both $f_0$ and $T$, in such a way that the condition $v_0^2=f_0^2/(m\gamma)^2 \gg T/m$ is always satisfied. This guarantees that, even for large $\gamma$, the active force motion is not overwhelmed by thermal fluctuations (whose amplitude increases as $\gamma$ grows), and, thus, that the active component of the motion always plays a crucial role.
In this case, we observe that the parameters $1/\gamma$ and $\tau$ roughly play the same dynamical role since their independent increases lead to smoother trajectories, through which each particle is able to cover longer distances before the diffusive regime is approached.
The velocity change is determined by three contributions: the thermal acceleration, the active force, and the Stokes term.
Its evolution is controlled by the typical times, $\tau$ and $ 1/\gamma$, and, thus, by the larger between the persistence and the inertial time, in agreement with the observations of Fig.~\ref{fig:trajfree}.
When one of these two times is large the velocity changes slowly and, consequently, position trajectories become smooth.

\subsection{Steady-state properties}

The dynamics~\eqref{eq:motion_free} can be exactly solved because of its linearity, and we are able to predict the steady-state probability distribution function and the correlation matrix, whose calculations are reported in Appendix~\ref{app:steady_states}.
In the well-known overdamped active regime in the absence of external forces, the self-propulsion induces an effective particle velocity, $\dot{\mathbf{x}}=\mathbf{v}$, while, in the underdamped regime, $\mathbf{v}$ and $\mathbf{f}_a$ represent different degrees of freedom. Interestingly, even in the simplest case of potential-free particles, a non-zero correlation between $\mathbf{v}$ and $\mathbf{f}_a$ appears, that is
\begin{equation}
\label{eq:vfacorr_free}
\langle \mathbf{v} \cdot \mathbf{f}_a \rangle=  2 \frac{f_0^2}{m \gamma} \frac{\tau\gamma}{1+\tau\gamma} \,.
\end{equation}
The diagonal elements of the correlation matrix esplicitly read:
\begin{flalign}
\label{eq:vvcorr_free}
\langle \mathbf{v}^2 \rangle&= 2\frac{T}{m} + 2\frac{f_0^2}{m^2 \gamma^2} \frac{\tau\gamma}{1+\tau\gamma} \,,\\
\label{eq:ffcorr_free}
\langle \mathbf{f}_a^2 \rangle&=2 f_0^2 \,.
\end{flalign}
As expected in the equilibrium limit of vanishing self-propulsion , $f_0\to0$, the velocity variance is enterely determined by the solvent temperature, namely $\langle \mathbf{v}^2 \rangle \to 2T/m$, the cross correlation $\langle \mathbf{v} \cdot \mathbf{f}_a \rangle$ vanishes, and, thus, the variable $\mathbf{f}_a$ becomes irrelevant for the particle dynamics.
This regime is analog to the well-known case of passive equilibrium colloids.
In the limit $\tau\gamma \ll 1$ (small persistence time with respect to the inertial time), the dynamics approach again the limiting case of equilibrium colloids. This is also clear considering Eq.~\eqref{eq:fa_free} and dropping the time derivative of the active force so that $\mathbf{f}_a\approx f_0\sqrt{2 \tau} \boldsymbol{\xi}$.
Instead, for $\tau\gamma \gg 1$, the active force becomes relevant and Eqs.\eqref{eq:vfacorr_free} and \eqref{eq:vvcorr_free}, approach the limiting values $\langle \mathbf{v} \cdot \mathbf{f}_a \rangle\approx 2 \frac{f_0^2}{m\gamma}$ and $\langle \mathbf{v}^2 \rangle \approx 2T/m+2f_0^2/(m\gamma)^2$, respectively.
Since the active force provides an additional contribution to the velocity variance,  the amplitude of the velocity fluctuations increases.

By the knowledge of the steady-state correlation matrix, we can derive
the steady-state probability distribution function, $p(\mathbf{v},\mathbf{f}_a)$, that is given by a multivariate Gaussian coupling $\mathbf{f}_a$ and $\mathbf{v}$. The distribution can be expressed as:
\begin{equation}
\label{eq:prob_free_tot}
p \left(\mathbf{v}, \mathbf{f}_a\right)=p \left(\mathbf{v} | \mathbf{f}_a\right) p \left( \mathbf{f}_a\right)
\end{equation}
where $p \left( \mathbf{f}_a\right)$ is the marginal distribution of the active force
\begin{equation}
\label{eq:pfa_marginal}
p \left( \mathbf{f}_a\right)\propto \exp{\left(  -\frac{\mathbf{f}_a^2}{2f_0^2}   \right)} \,,
\end{equation}
and $p \left(\mathbf{v} | \mathbf{f}_a\right)$ is the conditonal probability distribution of the velocity at fixed self-propulsion:
\begin{equation}
\label{eq:prob_free_2}
p(\mathbf{v}| \mathbf{f}_a )\propto \exp{\left( -\frac{\beta}{2} m\biggl[ \mathbf{v} - \mathbf{u} \biggr]^2 -\frac{\mathbf{f}_a^2}{2f_0^2}   \right)} \,.
\end{equation}
The term $\mathbf{u}=\left\langle \mathbf{v}\left(\mathbf{f}_a\right)\right\rangle$ represents the average value assumed by $\mathbf{v}$ upon fixing $\mathbf{f}_a$, that is:
\begin{equation}
\label{eq:u_free}
\mathbf{u}= - \frac{\tau}{1+\tau\gamma} \frac{{\mathbf{f}_a}}{m} \,.
\end{equation}
The coefficient $\beta$ plays the role of an inverse kinetic temperature and reads
\begin{flalign}
\beta=\frac{1}{ T }  \left( 1+\frac{f_0^2}{m T}\frac{\tau}{\gamma} \frac{\tau\gamma}{\left(1+\tau\gamma\right)^2}  \right)^{-1} \,.
\end{flalign}
Remarkably, $\mathbf{u}$ vanishes in the limits $\tau\to0$  and $\gamma\to\infty$  where the system approaches two well-known equilibrium regimes.
In the former case, the system displays an underdamped passive behavior at the equilibrium temperature $T$ while, in the latter case, the system is characterized by an overdamped active motion.
In the athermal limit, $T\to0$, the shape of the distribution remains unchanged with reduced velocity fluctuations controlled by $\beta \to \beta_0$
$$
\beta_0=m \frac{\left(1+\gamma\tau\right)^2}{f_0^2 \tau^2} \,,
$$
consistent with the well-known result for the overdamped AOUP. 
Indeed, for $\tau\gamma\gg 1$ and  $\tau\gamma\ll 1$, the velocity fluctuations are given by $f_0^2/(m\gamma)^2$ and $f_0^2\tau^2/m^2$, respectively.

\subsection{Time correlation of the velocity and mean square displacement}
\begin{figure}[t]
\centering
\includegraphics[width=0.9\columnwidth,keepaspectratio]
{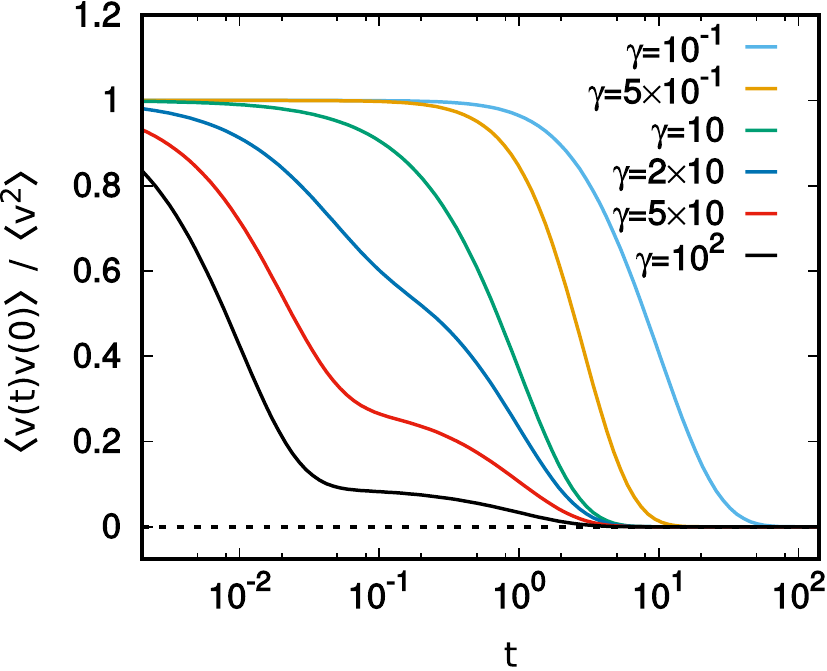}
\caption{\label{fig:vv_free}
Temporal velocity correlation normalized with its variance, i.e. $\langle \mathbf{v}(t) \mathbf{v}(0) \rangle/\langle \mathbf{v}^2 \rangle$, as a function of $t$, for several values of $\gamma$.
The other parameters are $\tau=1$, $T=10^{-3}$ and $f_0=1$.
}
\end{figure}
As shown in the Appendix~\ref{appendixB}, we can calculate the steady-state velocity correlation that is a combinations of time exponentials:
\begin{equation}
\begin{aligned}
\label{eq:vtv0corr_free}
\langle \mathbf{v}(t) \cdot\mathbf{v}(0)\rangle&=2\frac{T}{m} e^{-\gamma t} + 2\frac{f_0^2}{m^2 \gamma^2}\frac{1}{1-\tau^2\gamma^2}\times\\
&\times\left[\tau\gamma e^{-\gamma t} -\tau^2\gamma^2 e^{-t/\tau} \right] \,.
\end{aligned}
\end{equation}
At variance with the case of overdamped active particles,  $\langle \mathbf{v}(t) \cdot\mathbf{v}(0)\rangle$ decays with two typical times, $1/\gamma$ and $\tau$.
The first term in Eq.~\eqref{eq:vtv0corr_free} represents the equilibrium part of the correlation which does not vanish  in the equilibrium limits, namely $f_0\to0$ or $\tau\to0$.
The second term is due to the joint action of self-propulsion and inertia and is positive for every choice of $\gamma$ and $\tau$.

For small values of $\tau\gamma$, the decay is dominated by the inertial term because the amplitude of $e^{-t/\tau}$ is smaller, while, for large $\tau\gamma$, the time decay of the correlation is mainly determined by the term $e^{-t/\tau}$ (second term of the square bracket).
Fig.~\ref{fig:vv_free} shows several time-correlations, $\langle \mathbf{v}(t) \mathbf{v}(0) \rangle/\langle \mathbf{v}^2 \rangle$, normalized with their variances (given by Eq.~\eqref{eq:vvcorr_free}) for several values of $\gamma$ and fixed $\tau=1$.
In the overdamped regime (such that $\gamma \gg 1/\tau$), the velocity correlation displays a double decay: after initial decay occurring for $t \approx 1/\gamma$, a slow decay controlled by $\tau$ is detected.
Decreasing $\gamma$ the distinction between the two time-regimes becomes less pronounced till to disappear in the underdamped regime, $\gamma \ll 1/\tau$, where  $\langle \mathbf{v}(t) \mathbf{v}(0) \rangle$ starts decaying after $1/\gamma$.

We can easily calculate the mean-square-displacement, $\text{MSD}(t)=\langle \Delta x(t)^2 \rangle$, by integrating the stationary velocity correlation (Eq.~\eqref{eq:vtv0corr_free}):
\begin{equation}
\label{eq:MSD}
\begin{aligned}
&\text{MSD}(t)=\int^t_0 dt' \int^{t'}_0 dt'' \langle \mathbf{v}(t')\cdot\mathbf{v}(t'')\rangle \\
&=2\left( \frac{T}{m\gamma^2} - \frac{f_0^2 \tau}{m^2\gamma^3 (\gamma^2\tau^2-1)}  \right) \left[ t\gamma + e^{-t \gamma} -1  \right]\\
&+\frac{2}{m^2}\frac{f_0^2 \tau^4}{\gamma^2\tau^2-1} \left[ \frac{t}{\tau}+ e^{-t/\tau} -1 \right]
\end{aligned}
\end{equation}
where $t'>t''$.
The active force, on the one hand, introduces the typical time $\tau$ ruling the passage from a ballistic to a diffusive regime (third line), on the other hand, changes the amplitude of the equilibrium term  (second line).
In the small-time regime, when both $t/\tau \ll 1$ and $\gamma t \ll 1$, the mean square displacement displays a full ballistic regime, that is:
\begin{equation}
\lim_{t\ll \tau ; \, t \ll 1/\gamma} \text{MSD}(t) \approx  \left( \frac{T}{m} + \frac{f_0^2}{m^2}\frac{1}{1+\tau\gamma} \frac{\tau}{\gamma} \right) t^2 \,,
\end{equation}
simply obtained expanding Eq.~\eqref{eq:MSD} in powers of $t/\tau$ and $t\gamma$ at the lower orders.
In this regime, the $\text{MSD}(t)$ is the sum of the usual thermal contribution (that is usually not easily experimentally accessible for equilibrium colloids) and a ballistic contribution due to the active force $\propto f_0^2$.
While the first term is $\gamma$ independent, the second displays a pronounced decreasing dependence on $\gamma$, which, in particular, becomes $1/\gamma^2$ in the regime $\gamma\tau\gg1$.
As expected, both in the small persistence and overdamped regimes, the amplitude of the self-propulsion should be very large to get a non-negligible contribution.
Instead, in the large persistence regime, in particular, when $\tau\gamma \gg 1$, the $\tau$-dependence disappears and the active force gives its maximal contribution that becomes larger and larger when $\gamma$ is decreased.
Instead, when $t/\tau \gg 1$ and $\gamma t \gg1$, i.e. in the large time regime, the $\text{MSD}(t)$ approaches a diffusive behavior, such that:
\begin{equation}
\begin{aligned}
\lim_{t\gg \tau ; \, t \gg 1/\gamma}\text{MSD}(t)& \approx 2\left(D_{t}+D_a \right) t \\
&= 2\left(\frac{T}{m \gamma} + \frac{f_0^2}{m^2} \frac{\tau}{\gamma^2} \right) t \,,
\end{aligned}
\end{equation}
where the effective diffusion coefficient, $D_{\text{eff}}=D_a+D_t$, is the sum of $D_t=\frac{T}{m \gamma}$, the thermal diffusion coefficient of a colloidal particle moving in a solvent and $D_a= \frac{f_0^2}{m^2} \frac{\tau}{\gamma^2}$, the active diffusion coefficient due to the self-propulsion.
 $D_{\text{eff}}$ decreases as $\gamma$ increases because the particle motion is hindered by the viscous force.
Moreover, the peculiarity of the active diffusion relies on the anomalous $\gamma$-scaling, being $D_a \propto 1/\gamma^2$ at variance with $D_t\propto 1/\gamma$.
Thus, at fixed $f_0$ and $T$ the overdamped regime ($\gamma\gg 1$) favors the thermal diffusion, while, in the underdamped regime ($\gamma \ll 1$), the term $D_a$ increases faster than $D_t$.
The term $D_a$ increases also as $\tau$ grows in agreement with the well-known result obtained in the overdamped regime.


\section{The harmonic potential}
\label{Sec:Harmonic}

\begin{figure}[t]
\centering
\includegraphics[width=0.95\columnwidth,keepaspectratio]
{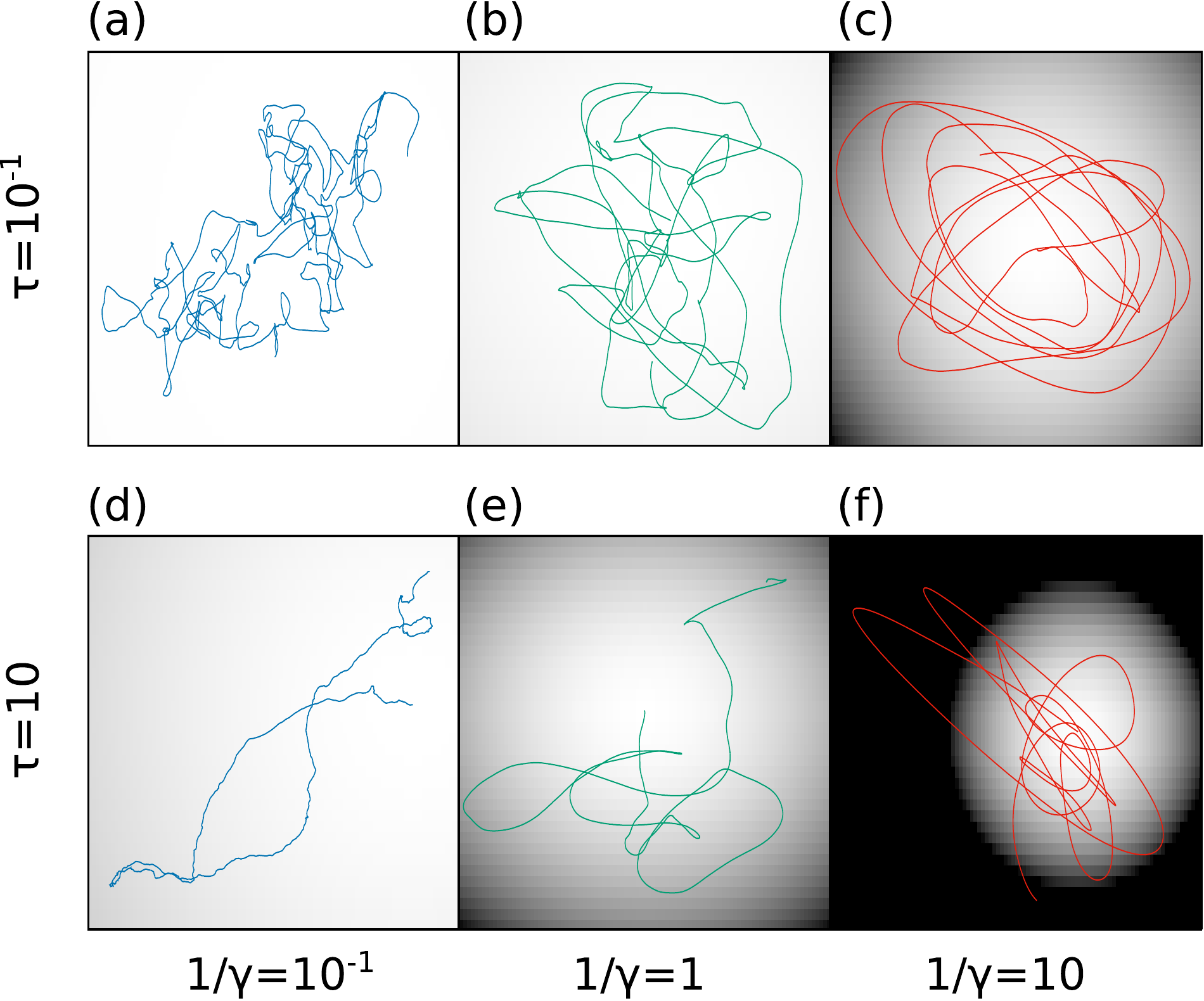}
\caption{\label{fig:trajharmonic}. 
Single-particle trajectories in the plane $xy$ evolving with the confining dynamics~\eqref{eq:motion_harmonic} up to the final time $\mathcal{T}=10^2$.
Panels (a), (b) and (c) (upper panels) are obtained with $\tau=10^{-1}$ while panels (d), (e) and (f) (bottom panels) with $\tau=10$.
Instead the couples [(a),(c)], [(b), (d)] and [(c), (f)] have been realized with $1/\gamma=10^{-1}, 1, 10$, respectively.
The trajectories are colored according to the value of $1/\gamma$.
The colored map (a scale of gray) plots the harmonic potential $U(\mathbf{x})=k \mathbf{x}^2/2$: the white and black colors correspond to zero and potential values larger than 10, respectively. 
The trajectories are obtained with $T=10^{-3}$ and $f_0=1$.
}
\end{figure}

The possibility of trapping a self-propelled particle using a confining force, which is approximatively harmonic, has been recently realized employing acoustic traps in the case of Janus particles~\cite{takatori2016acoustic} or using parabolic dishes in the case of Hexbug toy robots~\cite{dauchot2019dynamics}.
Several studies have been focused on the overdamped active regime considering both AOUP~\cite{szamel2014self, caprini2019activity, woillez2020active} and ABP dynamics \cite{hennes2014self, basu2019long, hoell2019multi, rana2019tuning, pototsky2012active, malakar2020steady} and  in the harmonic case a comparison has been made~\cite{caprini2019comparative, das2018confined}.
Here, we evaluate the influence of inertial forces on the dynamics of a self-propelled particle confined through a harmonic potential $U(\mathbf{x})=(k/2)\, \mathbf{x}^2$, where $k$ determines the potential strength.
In this case, the dynamics reads:
\begin{subequations}
\label{eq:motion_harmonic}
\begin{align}
\dot{\mathbf{x}}&=\mathbf{v} \,,\\
\dot{\mathbf{v}}& = - \gamma \mathbf{v} - \omega_0^2\,\mathbf{x}+ \frac{\mathbf{f}_a}{m} + \sqrt{2 \gamma \frac{T}{m}} \,\boldsymbol{\eta}\,,\\
\tau\dot{\mathbf{f}}_a &= -\mathbf{ f}_a + f_0\sqrt{2 \tau} \boldsymbol{\xi} \,
\end{align}
\end{subequations}
with $\omega_0^2=k/m$.
At variance with the potential-free case, here, three different typical times govern the dynamics.
In addition to the inertial time, $1/\gamma$, and the persistence time, $\tau$, the relaxation towards the minimum of the potential is determined by the typical time $\gamma/\omega_0^2$.
While the potential-free dynamics are simply controlled by the ratio between persistence and inertial time, $\tau\gamma$, here, a second ratio comes into play, that is $\tau \omega_0^2/\gamma$, i.e. the ratio between persistence and potential time.

The single-particle trajectories are reported in Fig.~\ref{fig:trajharmonic} for several values of $1/\gamma$ and $\tau$, keeping fixed the strength of the potential $k=1$ and superimposing the potential $U(\mathbf{x})=k\mathbf{x}^2/2$ as a scale of gray (white at the minimum of the potential and darker gray as the value of $U(\mathbf{x})$ increases).
For the smaller values of $\tau$ and $1/\gamma$, the dynamics behaves as in the case of overdamped passive colloids, i.e. shows fluctuations around the minimum of the potential, while for larger persistence time, $\tau$, as in the case of overdamped active motion, the trajectories explore spatial regions far from the minimum of the potential.
In analogy to the potential-free case, the increase of $1/\gamma$ leads to smother trajectories independently of their persistence time (see the sequences (a), (b) and (c) obtained with $\tau=10^{-1}$ and (d), (e) and (f) with $\tau=10$ ).
In particular, in the underdamped regime, for $1/\gamma =10$ (panels (c) and (f)), the particle performs almost circular trajectories that are squeezed when $\tau$ is increased.
This kind of behavior has not an overamped active counterpart and is entirely due to the interplay between inertia and self-propulsion. 
Finally, we observe that, as both $1/\gamma$ and $\tau$ are increased, particles have the capacity to explore larger regions of space.
As we will see later on, this is due to the growth of the positional variance of the distribution.

\subsection{Correlation matrix and steady-state probability distribution}
Hereafter, we present the exact result for the correlation matrix (see Appendix~\ref{app:steady_states}).
Since the self-propulsion is not influenced by the remaining degrees of freedom, the variance of the active force, $\langle \mathbf{f}_a^2\rangle$, is still determined by Eq.~\eqref{eq:ffcorr_free}.
In the harmonic potential case, particle velocity and position are uncorrelated, $\langle\mathbf{x}\cdot\mathbf{v}\rangle=0$, while both $\mathbf{v}$ and $\mathbf{x}$ develop a correlation with the self-propulsion:
\begin{flalign}
&\langle \mathbf{v}\cdot \mathbf{f}_a\rangle =  \frac{2}{m}\frac{f_0^2 \tau}{1+\tau\gamma+\tau^2 \omega_0^2} \,,\\
&\langle \mathbf{x}\cdot \mathbf{f}_a\rangle =  -\tau\langle \mathbf{v}\cdot \mathbf{f}_a\rangle \,.
\end{flalign}
Remarkably, the harmonic confinement term decreases the cross-correlation $\langle \mathbf{v}\cdot \mathbf{f}_a\rangle$ with respect to the potential-free case and introduces an anti-correlation between position and self-propulsion that is proportional to $\langle \mathbf{v}\cdot \mathbf{f}_a\rangle$ and increases as $\tau$ grows
 reaching a plateau for $\tau \to \infty$.
This occurs because a correlation between the elastic force, $-k \mathbf{x}$, and the self-propulsion naturally arises from the balance between $\mathbf{f}_a$ and the elastic force:  when the particle climbs on the potential it remains almost stuck at positions $ k \mathbf{x} \approx \mathbf{f}_a$ for a time $\sim \tau$.

The variances of particle position and velocity, i.e. the diagonal element of the correlation matrix, read:
\begin{flalign}
\label{eq:xx_corr_harmonic}
\langle \mathbf{x}^2\rangle &= 2\frac{T}{m \omega_0^2} + 2\frac{f_0^2\tau}{ m^2\omega_0^2 \gamma}\frac{1+\tau\gamma}{1+\tau\gamma+ \omega_0^2\tau^2 } \,,\\
\label{eq:vv_corr_harmonic}
\langle \mathbf{v}^2\rangle &= 2\frac{T}{m} +  2\frac{f_0^2 \tau}{ m^2\gamma}\frac{1}{1+\tau\gamma+ \omega_0^2\tau^2  } \,.
\end{flalign}
We observe that, in the equilibrium limit $f_0\to0$, the correlation matrix reduces to the well-known result of passive colloids in the underdamped regime: the cross-correlations vanish, $\langle \mathbf{x}^2 \rangle $ decreases as $k$ while $\langle \mathbf{v}^2 \rangle$ is $k$-independent. 
The same result can be achieved taking the limit $\tau\to0$ or $\gamma\to\infty$, in analogy with the case of potential-free particle.
The new aspect of the dynamics occurs in the large persistence regime where the self-propulsion is dominant (consider the athermal limit, $T=0$, for simplicity) in such a way that Eqs.~\eqref{eq:xx_corr_harmonic} and~\eqref{eq:vv_corr_harmonic} are controlled by the second terms $\propto f_0^2$.
Here, we can distinguish between two different cases depending on the ratio between the three typical times.
When $\tau\gamma \gg \tau k^2 \gg 1$, the self-propulsion affects the velocity variance through an additional term, that is the square of the swim velocity $v_0=f_0/\gamma$:  in this regime, the only average effect of the self-propulsion is a change of the kinetic temperature. Moreover, $\langle \mathbf{x}^2\rangle$ increases as $v_0^2 \tau/\gamma$, in agreement with the qualitative observations of Fig.~\ref{fig:trajharmonic}.
More interesting is the regime $\tau k^2 \gg \tau\gamma \gg 1$ (i.e. when the potential-time is smaller than the persistence one).
In this regime, the velocity variance reduces to
$$
\langle \mathbf{v}^2\rangle \approx 2 v_0^2\frac{\gamma}{\tau  \omega_0^2} \,,
$$
showing that particles slow down as $k$ grows, being $\omega_0^2=k/m$.
Correspondingly, $\langle \mathbf{x}^2\rangle \propto 1/k^2$, meaning that the typical equilibrium scaling with $k$ is changed with respect to passive particles.

The steady-state probability distribution, $p(\mathbf{x}, \mathbf{v}, \mathbf{f}_a)$, is a multivariate Gaussian involving $\mathbf{x}$, $\mathbf{v}$ and $\mathbf{f}_a$ and is expressed in terms of conditional probability distributions:
\begin{equation}
\label{eq:prob_harm_tot}
p \left(\mathbf{x}, \mathbf{v}, \mathbf{f}_a\right) = p(\mathbf{x}| \mathbf{v}, \mathbf{f}_a) \, p(\mathbf{v}|\mathbf{f}_a) \, p(\mathbf{f}_a) \,.
\end{equation}
The marginal distribution of $\mathbf{f}_a$ is still given by Eq.~\eqref{eq:pfa_marginal} while $p(\mathbf{x}| \mathbf{v}, \mathbf{f}_a)$ and $p(\mathbf{v}|\mathbf{f}_a)$ are:
\begin{flalign}
\label{eq:prob_harm_1}
&p(\mathbf{v}|\mathbf{f}_a) \propto \exp{\left(-\frac{\beta}{2} m\left( \mathbf{v} - \mathbf{u} \right)^2\right)}\\
\label{eq:prob_harm_2}
&p(\mathbf{x}| \mathbf{v}, \mathbf{f}_a) \propto \exp{\left( - \frac{\alpha}{2} \left( \mathbf{x} -  \mathbf{r}\right)^2 \right)} \,,
\end{flalign}
where we have introduced the symbols $ \mathbf{u}= \left\langle \mathbf{v}(\mathbf{f}_a) \right\rangle$ and $\mathbf{r}= \left\langle \mathbf{x}(\mathbf{f}_a, \mathbf{v}) \right\rangle$ to denote the conditional averages of velocity and position, obtained at fixed values of $\mathbf{f}_a$ and $\mathbf{f}_a$, $\mathbf{v}$, respectively: 
\begin{flalign}
\label{eq:u_harm}
 \mathbf{u} &=\tau(1+\tau\gamma\Gamma)\left\{\left[\frac{f_0^2\tau}{m T\gamma}+1+\tau\gamma\Gamma\right]^2+\frac{f_0^2\gamma\tau^3}{m T}\right\}\nonumber\\
&\times\frac{\left[ \frac{f_0^2\tau}{m T\gamma}\left( 1+ \omega_0^2 \tau^2 \right)+\left(1
+\tau\gamma\Gamma\right)^2  \right]^{-1}}{\left[  \frac{f_0^2\tau}{m T\gamma}\left( \left[1+\tau\gamma\right]^2+ \omega_0^2\tau^2 \right)+\left(1
+\tau\gamma\Gamma\right)^2  \right]} \frac{ \mathbf{f}_a}{m} \,,  \\
\label{eq:r_harm}
 \mathbf{r} &=  \frac{f_0^2 \tau^3}{m T} \frac{\mathbf{v}}{\left( 1+\tau\gamma\Gamma \right)^2 + \frac{f_0^2\tau}{m T\gamma} \left[ 1+ \omega_0^2\tau^2 \right]}  \nonumber\\
&-  \frac{1+\tau\gamma\Gamma+\frac{f_0^2\tau}{m T\gamma}}{\left( 1+\tau\gamma\Gamma \right)^2 + \frac{f_0^2\tau}{m T\gamma} \left[ 1+ \omega_0^2\tau^2 \right]} \tau^2 \frac{ \mathbf{f}_a}{m}  \,.
\end{flalign}
As in the case of overdamped active particles (AOUP), we have introduced the symbol 
$$
\Gamma=1+\frac{\tau}{\gamma}  \omega_0^2 \,,
$$
namely the effective friction due to the action of self-propulsion in harmonic potentials that increases as $k$ grows~\cite{caprini2019activity}.
Finally, the coefficients, $\alpha$, $\beta$, controlling the amplitude of the of velocity and position fluctuations are:
\begin{flalign}
\beta &= \frac{1}{T} \frac{ \left(1+\tau\gamma\Gamma \right)^2 + \frac{f_0^2 \tau}{m T\gamma}\left[ (1+\tau\gamma)^2 + \omega_0^2\tau^2 \right]}{\left[\left(1+\tau\gamma \Gamma\right)+\frac{f_0^2\tau}{m T\gamma}\right]^2 + \frac{f_0^2\tau^3\gamma}{m T}}\\
\alpha &= \frac{m \omega_0^2}{T} \frac{1+\tau\gamma\Gamma + \frac{f_0^2\tau}{m \gamma T}\left( 1+ \omega_0^2\tau^2 \right)}{\left[\left(1+\tau\gamma \Gamma\right)+\frac{f_0^2\tau}{m T\gamma}\right]^2 + \frac{f_0^2\tau^3\gamma}{mT}} \,.
\end{flalign}
We observe that both $\beta$ and $\alpha$ approach to the equilibrium values $1/T$ and $m  \omega_0^2/T$ for $\tau\to0$, as expected.
Instead, the limit $f_0\to 0$ should be performed carefully because of the divergence of Eq.~\eqref{eq:pfa_marginal}: one needs to integrate over $\mathbf{f}_a$, before taking this limit and getting the correct result.

The athermal limit, $T\to0$, reveals the leading contribution due to the presence of the self-propulsion that simplifies the previous expressions and illustrates the role of the self-propulsion more clearly:
\begin{flalign}
 \mathbf{u} &= \frac{\mathbf{f}_a}{m} \frac{\tau}{\left(1+ \omega_0^2\tau^2\right)\left( 1+\tau\gamma\Gamma \right)}  \\
 \mathbf{r} &=  \frac{\tau^2 }{1+ \omega_0^2\tau^2} \left( \gamma \mathbf{v}-\frac{\mathbf{f}_a}{m} \right) \,.
\end{flalign}
Remarkably, $ \mathbf{u}$ is proportional to $\mathbf{f}_a$: the larger is the the self-propulsion the larger is $ \mathbf{u}$.  Instead, the mean particle position is determined by the balance between $\gamma\mathbf{v}$ and $\mathbf{f}_a$.
We observe that the confinement reduces the value of both $ \mathbf{u}$ and $ \mathbf{r}$ slowing down the particle dynamics and pushing the particle towards the minimum of the potential.
In this limit, the increase of $\tau$ maximizes $ |\mathbf{r}|$ but leads to a vanishing $\mathbf{v}$, in such a way that ${\bf r} \approx - \mathbf{f}_a/m \omega_0^2$.

\subsection{Time correlations of position and velocity}

\begin{figure*}[t]
\centering
\includegraphics[width=0.95\linewidth,keepaspectratio]
{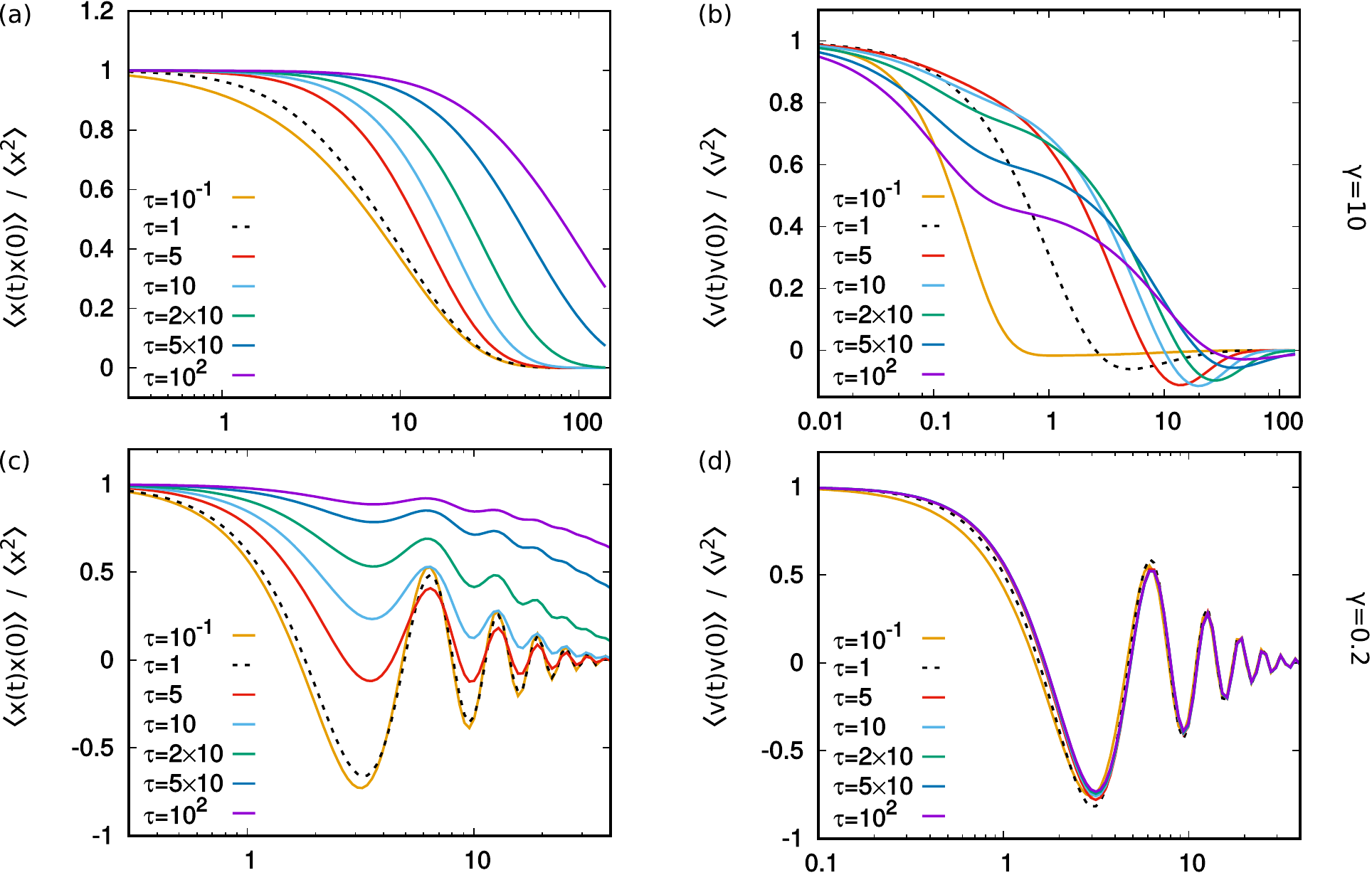}
\caption{\label{fig:corr_harmo}
Time correlations normalized with their variances of a harmonically confined self-propelled particle.
Panels (a) and (c) show $\langle \mathbf{x}(t) \mathbf{x}(0) \rangle/\langle \mathbf{x^2}\rangle$ for $\gamma=10, 0.2$, respectively, while panels (b) and (d)  show $\langle \mathbf{v}(t) \mathbf{v}(0) \rangle/\langle \mathbf{v^2}\rangle$ for $\gamma=10, 0.2$, respectively. 
Each panel plots the curves with different values of $\tau$, according to the legend.
The other parameters are $T=10^{-3}$ and $f_0=1$.
}
\end{figure*}

As in the case of potential-free particles, we can analytically calculate the steady-state temporal correlations, as shown in Appendix~\ref{appendixC}.
In this case, the positional time-correlation, $\langle \mathbf{x}(t) \cdot \mathbf{x}(0)\rangle$, can be split into two terms: 
\begin{equation}
\label{eq:xxcorr_time}
\langle \mathbf{x}(t) \mathbf{x}(0)\rangle = \langle \mathbf{x}(t) \mathbf{x}(0)\rangle_T + \langle \mathbf{x}(t) \mathbf{x}(0)\rangle_A \,,  
\end{equation}
where $\langle x(t) x(0)\rangle_T$ is the thermal contribution of the correlation:
\begin{equation}
\label{eq:xxcorr_time_T}
\langle \mathbf{x}(t) \mathbf{x}(0)\rangle_T = 2\frac{T}{m  \omega_0^2} e^{-\gamma t/2 } \left[ \cos{\left(\omega t \right)} + \gamma \frac{\sin{\left(\omega t \right)}}{2 \omega}  \right] \,,
\end{equation}
where $\omega$ is defined as
\begin{equation}
\omega^2=  \omega_0^2-\frac{\gamma^2}{4} \,.
\end{equation}
The contribution~\eqref{eq:xxcorr_time_T} survives even in the absence of self-propulsion, $f_0 \to 0$, and corresponds to the well-known correlation of the particle position in the passive underdamped regime. Depending on the sign of $\omega^2$ (which is determined by the ratio between the inertial and the potential times) the system shows oscillations during its exponential decay. 
Instead, the second term in Eq.~\eqref{eq:xxcorr_time}, namely $\langle x(t) x(0)\rangle_A$, is entirely due to the active force and reads:
\begin{equation}
\begin{aligned}
\label{eq:xxcorr_time_A}
&\langle \mathbf{x}(t)\mathbf{x}(0)\rangle_A = 2\frac{f_0^2 \tau}{m^2\gamma  \omega_0^2} \frac{1}{(1+\tau^2  \omega_0^2)^2-\gamma^2\tau^2}\times\\
&\times\biggl[ \tau^3 \gamma  \omega_0^2 e^{-t/\tau} + e^{-\gamma t/2 }\left(   1-\tau^2\gamma^2+ \omega_0^2\tau^2\right) \cos(\omega t) \\
& \qquad\qquad+e^{-\gamma t /2 }\frac{\gamma}{2\omega} \left( 1-\tau^2\gamma^2+3\tau^2  \omega_0^2 \right)\sin{(\omega t)}\biggr]\,.
\end{aligned}
\end{equation}
This term survives also in the athermal regime, $T\to0$, and includes the whole effect of the active force.
In particular, $f_0$ is a simple multiplicative factor while $\tau$ has a more complex role.
On the one hand, $\tau$ introduces an additional time behavior (first term of the square bracket) with respect to the equilibrium prediction, Eq.~\eqref{eq:xxcorr_time}, on the other hand, affects the amplitude of the decay due to inertial effects (second and third terms in the square bracket).
The time-correlation $\langle \mathbf{x}(t) \mathbf{x}(0)\rangle/\langle \mathbf{x}^2\rangle$ is reported in Fig.~\ref{fig:corr_harmo}~(a) and~(c) for $\gamma=10, 0.2$, respectively, to show typical overdamped and underdamped cases.
For each case, we explore different values of $\tau$ to evaluate both large and small persistence regimes.
In the overdamped case (such that $\omega^2$ is real), the decay of $\langle \mathbf{x}(t) \mathbf{x}(0)\rangle$ is monotonic. For $\tau \lesssim 1/\gamma$ the curve collapse since the decay is only controlled by the inertial time.
Instead, in the regime $\tau \gtrsim 1/\gamma$, the decay is controlled by $\tau$ and, thus, becomes slower as the persistence is increased.
A more intriguing scenario occurs in the underdamped regime (such that $\omega^2$ is imaginary), where the decay of $\langle \mathbf{x}(t) \mathbf{x}(0)\rangle$ is characterized by many oscillations with decreasing amplitude.
Also in this case, for $\tau \lesssim 1/\gamma$ the curve collapse and are mainly determined by the thermal term of the correlation.
The increasing of $\tau$, on the one hand, reduces the amplitude of the oscillation, on the other hand, shifts the curves towards positive values, in such a way that for $\tau$ large enough $\langle \mathbf{x}(t) \mathbf{x}(0)\rangle$ does not approach negative values. Further increases of $\tau$ are even able to cancel the oscillations.

The velocity correlation, $\langle \mathbf{v}(t)\cdot \mathbf{v}(0) \rangle$, can be easily calculated from Eq.~\eqref{eq:xxcorr_time} because satisfies the following relation:
$$
\langle \mathbf{v}(t) \mathbf{v}(0)\rangle=- \frac{d^2}{dt^2} \langle \mathbf{x}(t) \mathbf{x}(0)\rangle \,.
$$
Thus, also the velocity correlation can be decomposed into a thermal, $\langle \mathbf{v}(t) \mathbf{v}(0)\rangle_T$, and an active contribution $\langle \mathbf{v}(t) \mathbf{v}(0)\rangle_A$:
\begin{flalign}
&\langle \mathbf{v}(t) \mathbf{v}(0)\rangle_T= 2 \frac{T}{m} e^{-\gamma t/2 } \left[ \cos{\left(\omega t \right)} - \gamma \frac{\sin{\left(\omega t \right)}}{2\omega}  \right] \,,
\\
&\langle \mathbf{v}(t) \mathbf{v}(0)\rangle_A=2\frac{f_0^2 \tau}{m^2 \gamma} \frac{1}{(1+\omega_0^2\tau^2 )^2-\gamma^2\tau^2}\times\\
&\times\biggl[ -\tau \gamma e^{-t/\tau} + e^{-\gamma t/2 }\left(1 + \omega_0^2\ \tau^2 \right) \cos(\omega t) \nonumber\\
& \qquad - \gamma e^{-\gamma t/2 }\frac{\left( 1-\omega_0^2\tau^2  \right)}{2\omega}\sin{(\omega t)}\biggr]\nonumber\,.
\end{flalign}
The velocity correlations are reported in Fig.~\ref{fig:corr_harmo}~(b) and~(d) for $\gamma=10, 0.2$, respectively, and show the same cases of $\langle \mathbf{x}(t) \mathbf{x}(0)\rangle$.
In the overdamped case ($\gamma=10$), $\langle \mathbf{v}(t) \mathbf{v}(0)\rangle$ collapse for $\tau \lesssim 1/\gamma$ (not shown). A first increase of $\tau$ leads the correlation to decay slower and perform an oscillation that lead $\langle \mathbf{v}(t) \mathbf{v}(0)\rangle$ to approach negative values.
The minimum of this oscillation moves towards larger times as $\tau$ is increased.
Moreover, for further values of $\tau$, $\langle \mathbf{v}(t) \mathbf{v}(0)\rangle$ shows a double decay regime, after a first fast decay controlled by $1/\gamma$ a second slow decay controlled by $\tau$ occurs.
The underdamped case ($\gamma=0.2$), displays a different scenario since all the curves are almost collapsed.

\section{Virial pressure}
\label{Sec:Virial}

A system of $N$ particles enclosed in a region by some walls, here described through a confining potential, exert on the walls a force whose average per unit surface value is the mechanical pressure.
In order to calculate explicitly the pressure, $P$, we use the virial theorem~\cite{widom2002statistical} which relates the so-called virial of the external (wall) forces ${\bf F} \cdot {\bf x}$ to the pressure through the formula:
$$
\frac{N}{d  L^d}\langle {\bf F}\cdot {\bf x}\rangle =-\, P \,
$$
where $d$ is the dimensionality, $L^d$ the volume of the enclosing region and $N$ the number of particles.
The virial can be obtained by the method of Falasco et al.~\cite{falasco2016mesoscopic,marconi2016pressure} and
in the case of confining harmonic walls,
a simple calculation yields the following  formula for the virial pressure~\cite{marini2017pressure}:
\begin{flalign}
&
P =N \frac{T}{L^2}\left(1+
 \frac{D_a}{D_t}  \left(
 \frac{1+\tau \gamma}{1 +\tau\gamma+\tau^2\omega_0^2}
  \right)     \right) \nonumber\\
 &=
 N\frac{T}{L^2}\left(1+
 \frac{ f_0^2}{m T} \frac{\tau}{\gamma}  \left(
 \frac{1+\tau \gamma}{1 +\tau\gamma+\tau^2\omega_0^2}\right)\right) \,,
\end{flalign}
where the first term is the standard osmotic pressure contribution whereas the second term represents the so-called swim
pressure~\cite{takatori2014swim}. 
Notice that the pressure is proportional to the variance of the velocity given by Eq.~\eqref{eq:vv_corr_harmonic}.
 In the limit $T\to0$, the osmotic pressure vanishes and the virial pressure is entirely determined by the $T$-independent
 swim pressure contribution. 

In the overdamped limit, $\tau \gamma\gg 1$,  the pressure becomes
$$
P\approx
 \frac{N}{L^2}\left(T+
 \frac{ f_0^2}{m } \frac{\tau}{\gamma}  \left(
 \frac{1}{1+\frac{\tau}{\gamma}\omega_0^2}\right)\right) \,
$$
where the second term has the same form as the one calculated in the overdamped case in Ref~\cite{marini2017pressure}. 
In this case, the swim pressure is an increasing function of $\tau$ that approaches its maximal value in the large persistence regime, $ \tau\gamma \gg 1$, where, in particular, reduces to $f_0^2 / (T m \omega_0^2)$.

Instead, in the underdamped regime where $\tau\gamma\ll 1$,
we have
\begin{flalign}
P\approx
 N\frac{T}{L^2}\left(1+
 \frac{ f_0^2}{m T} \frac{\tau}{\gamma}  \left(
 \frac{1}{1 +\tau^2\omega_0^2}\right)\right) \,
\end{flalign}
so that the swim pressure initially increases 
with $\tau$ if $\tau \omega_0 < 1$, but decreases for  values of  the persistence time larger than $1/\omega_0$.
Such a non-monotonicity of the pressure with the persistence time is due to the fact that
when the potential is very strong (large $\omega_0^2$) the particle is confined to a little region near its bottom
and the value of the velocity rapidly fluctuates around the zero. Correspondingly, the kinetic energy of the particle
remains very small and we observe only a little contribution to the swim pressure.


\section{Conclusion}
\label{Sec:Conclusion}

In this article, we have studied the role of the inertia on the dynamics of a self-propelled particle free to move into the solvent or confined in a harmonic trap using an
underdamped generalization of the Active Ornstein-Uhlenbeck particle model 
which successfully reproduces the phenomenology of self-propelled particles in the overdamped regime.
Some remarkable aspects of the underdamped AOUP are:
\begin{itemize}
\item[i)] The interplay between inertia and self-propulsion produces smoother trajectories as the persistence and inertial times are increased, both for harmonic confinement and potential-free cases.
\item[ii)] At variance with the underdamped ABP model, we found exact expressions for the correlation matrix of position and velocity and the steady-state probability distributions both for potential-free and harmonic potentials cases. In both cases, a marked correlation between particle velocity and self-propulsion spontaneously arises and, in the harmonic case, the elastic force (or, in other words, the particle position) correlates with both particle velocity and self-propulsion.
\item[iii)] We determined  the full expressions for the temporal properties of the system, studying, in particular, the different time-regimes of the mean-square displacement (unconfined system) and the shape of the position and velocity time-correlations both for potential-free and harmonically confined systems, unveiling the role of persistence time and viscosity.
\end{itemize}

We remark that this treatment is more general than the standard AOUP in the overdamped regime since the model is described by multiple degrees of freedom, i.e. the velocity and the position of the particles, and the dynamics includes two distinct stochastic energy sources provided by the thermal noise and the active force. 
These ingredients lead to the appearance of different time-scales and render even the simple oscillator model to look truly out of equilibrium.
We shall explore in detail this aspect in a future publication through a detailed energetic analysis in the framework of stochastic thermodynamics.

\subsection*{Acknowledgements}
LC and UMBM acknowledge support from the MIUR PRIN 2017 project 201798CZLJ.

\subsection*{Data availability}
The data that support the findings of this study are available from the corresponding author
upon reasonable request.

\appendix


\section{Probability distribution with linear interactions}\label{app:steady_states}
The potential-free and harmonically confined dynamics,  Eqs.~\eqref{eq:motion_free} and~\eqref{eq:motion_harmonic}, respectively, are of the form:
$$
\dot{\mathbf{w}} = - \mathcal{A} \cdot \mathbf{w} + \sqrt{2} \boldsymbol{\sigma} \cdot \boldsymbol{\eta} \,,
$$
where $A$ and $\boldsymbol{\sigma}$ are the drift and the noise matrices, respectively, and $\mathbf{w}$ a vector of dimensions $n$ being $n$ the number of variables involved in the dynamics.
In this special case (linear interactions only), the correlation matrix, $\mathcal{C}$, can be determined solving the following matricial relation:
\begin{equation}
\label{eq:Appendix1}
\mathcal{A} \cdot \mathcal{C}+\mathcal{C}\cdot \mathcal{A}^{T} = 2\mathcal{D} \,,
\end{equation}
where $\mathcal{A}^{T}$ is transpose matrix of $\mathcal{A}$ and $\mathcal{D}=\boldsymbol{\sigma}\cdot\boldsymbol{\sigma}^{T}$ the diffusion matrix.
In addition, the steady-state probability ditribution, $p(\mathbf{w})$, assumes the simple form:
$$
p(\mathbf{w})\propto \exp{\left( - \mathbf{w}\cdot \mathcal{C}^{-1} \cdot \mathbf{w}  \right)} \,,
$$
where $\mathcal{C}^{-1}$ is the inverse matrix of $\mathcal{C}$.

\subsection{Probability distribution functions in the potential-free case}

Applying the general method described above (Eq.~\eqref{eq:Appendix1}), where the vector $\mathbf{w}$ is formed by $\mathbf{v}$ and $\mathbf{f}_a$,
the distribution function in the potential-free case, $p(\mathbf{v}, \mathbf{f}_a)$, reads:
\begin{equation*}
p(\mathbf{x}, \mathbf{v}, \mathbf{f}_a) = \exp{\left( - \frac{\mathbf{v}^2}{2} C^{-1}_{\mathbf{v}\mathbf{v}} - \frac{\mathbf{f_a}^2}{2} C^{-1}_{\mathbf{f}_a\mathbf{f}_a}   - \mathbf{v f_a} C^{-1}_{\mathbf{v} \mathbf{f}_a} \right)} \,,
\end{equation*}
where 
\begin{flalign*}
C_{\mathbf{v}\mathbf{v}}^{-1}&=  \frac{m}{T} \frac{1}{\left(1+\frac{f_0^2\tau}{m\gamma T} \frac{1}{(1+\tau\gamma)^2}\right)}\\
C_{\mathbf{f}_a \mathbf{f}_a}^{-1}&=   \frac{1}{f_0^2} \frac{\left(1+\frac{f_0^2\tau}{m\gamma T} \frac{1}{1+\tau\gamma}\right)}{\left(1+\frac{f_0^2\tau}{m\gamma T} \frac{1}{(1+\tau\gamma)^2}\right)}   \\
C_{\mathbf{v} \mathbf{f}_a}^{-1}&=  - \frac{\tau}{T\left(1+\tau\gamma\right)}\frac{1}{\left(1+\frac{f_0^2\tau}{m\gamma T} \frac{1}{(1+\tau\gamma)^2}\right)}      
\end{flalign*}
Introducing the vector $\mathbf{u}=\left\langle \mathbf{v}(\mathbf{f}_a) \right\rangle$ given by Eq.~\eqref{eq:u_free}, we get the final result for the probability distribution that can be expressed as Eqs.~\eqref{eq:prob_free_tot},~\eqref{eq:pfa_marginal} and~\eqref{eq:prob_free_2}.

\subsection{Probability distribution function in the harmonic case}

In the harmonic case, the vector $\mathbf{w}$ is formed by $\mathbf{v}$, $\mathbf{x}$ and $\mathbf{f_a}$ and, thus, 
the distribution function, $p(\mathbf{x}, \mathbf{v}, \mathbf{f}_a)$, can be expressed as
\begin{equation*}
\begin{aligned}
p(\mathbf{x}, \mathbf{v}, \mathbf{f}_a) = &\exp{\left(- \frac{\mathbf{x}^2}{2} C^{-1}_{\mathbf{x}\mathbf{x}} - \frac{\mathbf{v}^2}{2} C^{-1}_{\mathbf{v}\mathbf{v}} - \frac{\mathbf{f_a}^2}{2} C^{-1}_{\mathbf{f}_a\mathbf{f}_a}\right)} \times\\
&\times\exp{\left(- \mathbf{xv} C^{-1}_{\mathbf{x}\mathbf{v}} - \mathbf{x f_a} C^{-1}_{\mathbf{x} \mathbf{f}_a} - \mathbf{v f_a} C^{-1}_{\mathbf{v} \mathbf{f}_a} \right)} \,,
\end{aligned}
\end{equation*}
where 
\begin{flalign*}
C_{\mathbf{x}\mathbf{x}}^{-1}&=\frac{m\omega_0^2}{T} \frac{1+\tau\gamma\Gamma + \frac{f_0^2\tau}{m\gamma T}\left( 1+\omega_0^2\tau^2 \right)}{\left[\left(1+\tau\gamma \Gamma\right)+\frac{f_0^2\tau}{m T\gamma}\right]^2 + \frac{f_0^2\tau^3\gamma}{m T}}\\
C_{\mathbf{v}\mathbf{v}}^{-1}&=\frac{m}{T} \frac{ \left(1+\tau\gamma\Gamma \right)^2 + \frac{f_0^2 \tau}{m T\gamma}\left[ (1+\tau\gamma)^2 +\omega_0^2 \tau^2 \right]}{\left[\left(1+\tau\gamma \Gamma\right)+\frac{f_0^2\tau}{m T\gamma}\right]^2 + \frac{f_0^2\tau^3\gamma}{m T}}\\
C_{\mathbf{f}_a\mathbf{f}_a}^{-1}&=\frac{1}{f_0^2}   \frac{\left[ 1+\tau\gamma \Gamma + \frac{f_0^2\tau}{m T\gamma}  \right] \left[ 1+\tau\gamma \Gamma + \frac{f_0^2\tau}{m T\gamma}\left(1+\tau\gamma  \right)  \right]  }{\left[\left(1+\tau\gamma \Gamma\right)+\frac{f_0^2\tau}{m T\gamma}\right]^2 + \frac{f_0^2\tau^3\gamma}{m T}}
\\
C_{\mathbf{x}\mathbf{v}}^{-1}&=-\frac{f_0^2 \omega_0^2 \tau^3}{T^2} \frac{1}{\left[\left(1+\tau\gamma \Gamma\right)+\frac{f_0^2\tau}{m T\gamma}\right]^2 + \frac{f_0^2\tau^3\gamma}{m T}}\\
C_{\mathbf{x}\mathbf{f}_a}^{-1}&=\frac{\omega_0^2\tau^2}{T} \frac{1+\tau\gamma\Gamma+\frac{f_0^2\tau}{T\gamma}}{\left[\left(1+\tau\gamma \Gamma\right)+\frac{f_0^2\tau}{T\gamma}\right]^2 + \frac{f_0^2\tau^3\gamma}{T}}
\\
C_{\mathbf{v}\mathbf{f}_a}^{-1}&=-\frac{\tau}{T} \frac{1+\tau\gamma\Gamma + \frac{\tau f_0^2}{m T\gamma}\left( 1+\tau\gamma \right)}{\left[\left(1+\tau\gamma \Gamma\right)+\frac{f_0^2\tau}{m T\gamma}\right]^2 + \frac{f_0^2\tau^3\gamma}{m T}}
\end{flalign*}
Defining $\mathbf{u}=\left\langle \mathbf{v}(\mathbf{f}_a) \right\rangle$ and $\mathbf{r}= \left\langle \mathbf{x}(\mathbf{f}_a, \mathbf{v}) \right\rangle$ as in Eqs.~\eqref{eq:u_harm} and~\eqref{eq:r_harm}, respectively, we obtain the final expression for the probability distribution function, Eqs.~\eqref{eq:prob_harm_tot},~\eqref{eq:prob_harm_1},~\eqref{eq:prob_harm_2} and~\eqref{eq:pfa_marginal}.

\section{Time correlations in the potential-free case}
\label{appendixB}

The inertial model is described by the two coupled equations for $x(t)$ and $f^a(t)$ (that we consider in one-dimension without loss of generality):
\begin{eqnarray}
&&
\label{falangevin}
\dot f^a(t) =-\frac{1}{\tau} f^a(t)+ f_0  \sqrt{\frac{2 } {\tau} }\xi(t)\\&&
\dot v(t) =  -\gamma v(t) +\frac{f^a(t)}{m}   + \sqrt{2 \gamma \frac{T}{m}} \eta(t)
\label{dynamicequation0}
\end{eqnarray}
%
%
A solution of the problem is
\begin{flalign}
\label{velocityequation} 
v(t)=v_0 e^{-\gamma t} +e^{-\gamma t}\int_0^t e^{\gamma t'} g(t')
\end{flalign}
where $v_0$ is the initial value of the velocity and the $g(t)$ represents the effective noise resulting form the 
combined action of the thermal noise and the active force:
\begin{flalign*}
\label{gfluctuation}
g(t)= \frac{f^a(t)}{m}   + \sqrt{2 \gamma \frac{T}{m}} \eta(t)  \, .
\end{flalign*}
It has the following averages:
\begin{flalign}
 &\langle g(t) \rangle =0 \,,\\
 &\langle g(t') g(t'')\rangle =2\gamma \frac{T}{m} \delta(t'-t'')+ \frac{f_0^2}{m^2} e^{-|t'-t''|/\tau} \,.
 \label{gaverage}
 \end{flalign}
In the steady-state, i.e. in the limit $t-t'\gg\tau$ and $t-t'\gg1/\gamma$, only the second term in Eq.~\eqref{velocityequation} 
contributes to the velocity correlation and we obtain:
\begin{flalign*}
&\langle v(t) v(0) \rangle= \frac{T}{m}  e^{-\gamma (t-t')} \\
&+\frac{ f_0^2}{m^2\gamma^2}    \frac{1} {1-\tau^2\gamma^2}   
 \left[ \tau\gamma e^{ -\gamma (t-t') }     -       \tau^2 \gamma^2 e^{-( t-t')/\tau}  \right] \nonumber
\end{flalign*}
That corresponds to Eq.~\eqref{eq:vtv0corr_free}.

\section{Time correlations for the case of harmonic potential}
\label{appendixC}

We cast the evolution equation for the position of the AOUP particle in the following form:
\begin{flalign}
\ddot x(t) =  -\gamma \dot x(t) -\frac{k }{m} x(t)  +\frac{f^a(t)}{m} + \sqrt{2 \gamma \frac{T}{m}} \eta(t)
\label{dynamicequationarm}
\end{flalign}
with $f^a$ given by Eq.~\eqref{falangevin}.
The particular solution corresponding to the Eq.~\eqref{dynamicequationarm} is:
\begin{flalign*}
x_{p}(t)=-e^{\lambda_{-} t} \int_0^t  dt' \frac{  g(t') \, e^{-\lambda_{-} t' }} { (\lambda_{+}-\lambda_{-})  } 
+e^{\lambda_{+} t} \int_0^t  dt' \frac{  g(t') \, e^{-\lambda_{+} t' }} { (\lambda_{+}-\lambda_{-})   } 
\end{flalign*}
%
with constants given by:
$$
\lambda_{\pm}=-\frac{\gamma\pm i \,2\omega}{2} \,,
$$
and
$$
\omega^2= \omega_0^2-\gamma^2/4 \,.
$$
Taking the noise averages and using \eqref{gaverage} we obtain the  correlation functions~\eqref{eq:xxcorr_time_T} and~\eqref{eq:xxcorr_time_A}.



\bibliographystyle{apsrev4-1}

\bibliography{under.bib}


\end{document}